\documentclass{article} 
\usepackage{aaai}

\newcommand{\ra}{\rightarrow}
\newcommand{\la}{\leftarrow}
\newcommand{\lra}{\leftrightarrow}

\begin{document}
\title{A  note on the Declarative reading(s) of Logic Programming}
\author{ Marc Denecker\\
Department of Computer Science, K.U.Leuven, \\
Celestijnenlaan 200A, B-3001 Heverlee, Belgium. \\
Phone: +32 16 327555 --- Fax:  +32 16 327996 \\
email: marcd@cs.kuleuven.ac.be}
\maketitle%

\begin{abstract}

  This paper analyses the declarative readings of logic programming.
  Logic programming - and negation as failure - has no unique
  declarative reading. One common view is that logic programming is
  a logic for default reasoning, a sub-formalism of default logic or
  autoepistemic logic. In this view, negation as failure is a modal
  operator. In an alternative view, a logic program is interpreted as
  a definition. In this view, negation as failure is classical
  objective negation. 

  From a commonsense point of view, there is definitely a difference
  between these views. Surprisingly though, both types of declarative
  readings lead to grosso modo the same model semantics. This note
  investigates the causes for this.

\end{abstract}

\section{Introduction}

Logic's fundamental role in the area of computing and artificial
intelligence, is its use for knowledge representation. There may be
innumerable ways in which some domain knowledge can be encoded in a
logic theory; however, there is one principle which most consider as
the canonical way of using logic.  {\em Declarative knowledge
representation} operates according to the following principle:
\begin{quote}
  the expert represents his domain knowledge by a set of
  formal statements that are {\em true} in the problem domain.
\end{quote}
This principle relies on the ability of the expert to interpret a
logical axiom as a clear and precise statement about the domain of
discourse. The ability of interpreting the formulas of a logic as
meaningful statements about the problem domain -given some
interpretation of the user-defined symbols- is the {\em declarative
reading} of the logic. It is based on a clear understanding of the
connectors and quantifiers and of how composed axioms combine
these meanings. 

A declarative semantics of the logic can be defined as a formal study
of its -intuitive- declarative reading. As such it should contain the
following parts:
\begin{itemize}
\item a clear account of some declarative reading of the formulas and
theories in a logic;
\item a mathematical characterisation of a formal semantics;
\item a justification why and how this semantics characterises this
declarative reading.
\end{itemize} 
In a sense, a declarative semantics should relate a logic to some part
of the human cognition and understanding. Note that a simple way of
defining a declarative semantics of a logic is by providing an
embedding in another logic with a well-established declarative
semantics. Such semantics are sometimes called {\em transformational
semantics}.

The declarative reading of logical axioms gives logic a decisive
advantage over other languages. It allows the expert to compare a
formal statement with his or her knowledge and to evaluate its {\em
truth}, without going through the painful process of explicitly
constructing the mathematical semantics of the axioms. This ideal is
reached most clearly in classical logic. So, our common expert
knowledge allows us to recognize the statement
\[ \forall x. person(x) \ra male(x) \lor female(x)\]
as {\em true} (given the obvious intended interpretation of the
predicate symbols), and the statement
\[ \forall x. male(x) \lra female(x)\]
as {\em false}. We can do so on the basis of our intuitive
understanding, without having to construct the semantics of these
sentences, i.e. their class of models.

It is a fact that to be able to represent knowledge in a declarative
way and to benefit the potential advantages of this type of knowledge
representation, a human expert must have acquired a deep and precise
understanding of the declarative reading of the logic that he is
using. Ambiguities and unclarities on the level of declarative reading
cause ambiguities and unclarities on the level of knowledge
representation. For this reason, studies of declarative reading are of
key importance in the development of knowledge representation
methodologies.


This paper is a study of the declarative reading(s) of logic
programming. The history of the declarative semantics of logic
programming is well-known. Originally, the picture was simple and
clear: the declarative reading of a Horn logic program was the
declarative reading of a set of classical logic implications. The
introduction of negation as failure blurred this simple view. On the
one hand, negation as failure derived conclusions with a strong
commonsense appeal and turned out to be very useful and natural in
many practical situations.  On the other hand, the negation as failure
inference rule was unsound with respect to the declarative reading of
a program as a set of classical implications. As expressed by
Przymusinski \cite{Przymusinski89}, ``{\em we really do not want
classical logic semantics for logic programs. .. We want the semantics
of a logic program to be determined more by its commonsense
meaning.}''.  All main semantic investigations since the end of the
seventies (least model semantics \cite{vanEmden76}, completion
\cite{Clark78}, perfect model semantics \cite{Apt88,Przymusinski89},
stable model semantics \cite{Gelfond88}, well-founded model semantics
\cite{VanGelder91}) attempted to formalise and explicitate this
commonsense meaning of logic programs. 

The question considered in this paper is to what extent the {\em
commonsense meaning} referred to by Przymusinski has been identified:
what is or are the declarative reading(s) of logic programming and
what are the corresponding meanings of its symbols $not, \la$.  What
are the semantics that correspond best to these declarative readings?

The analysis of the declarative reading of logic programming is
complicated by at least two factors. One complication is that
different formal semantics exist (in 2-, 3- and 4-valued versions). In
particular, stable and well-founded semantics are generally considered
as the main ones. However, there is a second complication which is
more subtle and much more dangerous. Let me try to pinpoint this
problem.

It is well-known that logic programming can be used for representing
many different sorts of knowledge \cite{Baral94a}: (inductive)
definitions, defaults, reflective knowledge of experts, etc.., both
under well-founded semantics and stable model semantics.  Remarkable
though is that the {\em same} logic program or rule in a logic program
can be used to represent knowledge with a decidedly distinct
commonsense flavor. For example, we might represent the definition
that {\em dead means not alive} by the rule:
\[   dead \la not\ alive \]
The interpretation of this rule as a definition of dead is
explicitated by the completion semantics \cite{Clark78}, Clark's
embedding of logic programming in classical logic. If the rule
defining $dead$ is the only rule with $dead$ in the head, the
completion will contain the equivalence
\[      dead \lra \neg alive \] 
Since stable and well-founded models are models of the completion,
this rule is satisfied also in these models.

On the other hand, consider the reflective knowledge of the
distrustful man who is unhappy if he does not know that his wife is
faithful to him. His reflective knowledge can be represented by the
same rule (modulo renaming) as in the above scenario:
\[   unhappy \la not\ wife\_faithful \]
Indeed, Gelfond's embedding of logic programs in autoepistemic logic
\cite{Gelfond87} maps this rule to the autoepistemic formula:
\[      unhappy \la \neg K wife\_faithful \]
which directly represents the knowledge of the distrustful man.

Gelfond's embedding defines a transformational declarative semantics
for logic programming. It formed the knowledge theoretical foundation
for stable model semantics \cite{Gelfond88}; therefore we are entitled
to assume that stable semantics is conceived to represent this sort of
knowledge. Note here that negation as failure is interpreted as a
modal operator and the implication operator as classical implication.
This is a common feature of all embeddings of logic programming in
autoepistemic logic (AEL) and in default logic (DL). Another important
embedding is Marek and Truszczynski's one  to DL \cite{Marek89}. It
maps he above rule to:
\[ \begin{array}{c} :\neg wife\_faithful\\\hline unhappy\end{array}\]
For an overview of different embeddings see \cite{Marek93b}).

From the commonsense point of view, there is a definite distinction
between both pieces of knowledge.
The definition of dead expresses that in the actual state of the
world, dead and alive are mutually exclusive: the exclusive ``or''
holds between them. If the expert does not know whether alive is true
or false, then he does not know whether dead is true or false. On the
other hand, the knowledge of the distrustful man does not imply any
relationship between the unhappiness of the distrustful man and the
loyalty of the spouse in the actual state of the world. It is
perfectly possible that in the actual state of the world, she is not
loyal but he does not know and he is happy.

Consider the logic program program $\{ p \la not\ q\}$. All main
semantics coincide for it; its formal semantics (i.e. the set of its
models) is the unique model $\{p\}$. Yet, as illustrated above,
Clark's embedding and Gelfond's embedding assign two different
commonsense meanings to this program: 
\begin{itemize}
\item In the completion, the program states that $q$ is false (because
  of the empty definition) and that $p$ is true iff $q$ is false.

  The model represents the state of the world in which $p$ is true and
  $q$ is false. According to the completion, this is the only possible
  state of the world.

\item The program in Gelfond's embedding states that $p$ holds if $q$
  is not known to be true.
  
  Since in this interpretation, the program has no knowledge about
  $q$, $q$ is unknown, hence $p$ is entailed. Note that contrary to
  the first reading, here $q$ is not known to be false.

  What is proven in \cite{Gelfond88}, is that the stable model
  represents the set of believed atoms of the above theory: $\{ p \}$
  means that $p$ is believed, and that the truth of $q$ is not
  believed.

  Based on our intuition, what can be said about the possible states
  of the world in this reading?  Obviously, since $q$ is unknown,
  there should be states in which $q$ is true and states in which $q$
  is false; $p$ should be true always. Hence, the two possible
  states are $\{p\}$ and $\{q,p\}$.

  Moore \cite{Moore84} defined a possible world semantics for
  AEL. Intuitively, a possible world model is a set of possible states
  of the world according to the expert's knowledge. The unique
  possible world model of the theory $\{ p\la \neg Kq\}$ is
  exactly the set $\{\{p\},\{q,p\}$; this  confirms our intuition.
\end{itemize}

The above discussion leads to a key point of this paper: even if we
know the models of a logic program, we still cannot decide the
intended declarative reading. We can only know -to some extent- what
is the declarative reading if we know also what is the role of the
model. A mathematical definition of some collection of {\em models} of
theories in a logic cannot {\em define} a declarative reading. In this
sense, a model theory is not a declarative semantics in its own right.

In the sequel, a model semantics will be called a {\em possible state
semantics} for some declarative reading if it characterises the
possible states of the world; a model theory will be called an {\em
atomic belief set semantics} of some declarative reading if it
characterises the sets of believed atoms in this declarative reading.

\vspace{5mm}

It is not a simple task to search for the declarative reading(s) of
logic programming. In the first place, many of the early semantical
studies in the context of logic programming are not primarily
concerned with finding and formalising a commonsense meaning, but are
more concerned with finding a mathematical justification for the reasoning
techniques in Prolog. 

Other studies are more focussed on the commonsense meaning, but fail
to give a clear account of the formalised declarative reading and the
role of the models. There is an enormous amount of mathematical
results on the relationships between different model
semantics. However, because atomic belief sets and possible states are
incomparable objects, it is a priori not clear what these relations
mean on the level of declarative reading. Moreover, in many knowledge
representation examples, one can observe that the same semantics is
used once as an atomic belief set semantics, once as an possible state
semantics.

In order to clarify the role of logic programming for Knowledge
representation, the question of declarative reading of logic
programming cannot be circumvented. In the rest of the paper I will
try to pinpoint the main ideas on declarative reading and the
confusions on this topic.

\section{Declarative readings of logic programming} 

An investigating of the transformational semantics that have been
proposed for logic programming, gives some insight in the possible
declarative readings of logic programming.

It seems to me that at least in the non-monotonic and A.I. oriented
part of the logic programming community, logic programming is now
routinely seen as a sub-logic of default logic or autoepistemic logic.
In this view, the negation as failure is interpreted as a modal
negation. This view has a natural motivation: a Prolog system is said
to infer a negative literal $not\ p$ when it is unable to prove $p$. A
natural way of modeling {\em failure to prove} in semantics is as {\em
not knowing}. From here, it was natural to interpret $not\ p$ as $\neg
K p$.

All main logic programming semantics - least model, supported model,
3-valued supported model, perfect model, stable model, 3-valued stable
model and well-founded semantics - have been justified as atomic
belief set semantics of diverse modal interpretations of logic
programming. The methods that have been used are analogous as Gelfond
and Lifschitz's justification for stable semantics: one defines an
embedding of logic programming to some non-monotone modal logic and the
models of the logic program are shown to be the set of believed
atoms. In these transformational semantics, models of logic
programming semantics systematically play the role of a set of
believed atoms. For an overview of these results, I refer to
\cite{Marek93b}.

On the other hand, I believe that there is also a persistent and
strong intuition that among all classical models of a logic program,
there is a canonical one (or at most a small number of canonical ones)
which represents the unique possible state of the world. The Clark
completion semantics was an early, weak attempt to identify this
canonical model. In this view, negation as failure does not need to be
interpreted as a modal operator: it is classical objective negation
denoting falsity in the canonical model. 

The main questions are what declarative reading of logic programming
could explain the existence of a unique possible state and how could
this unique model be mathematically characterized?

These questions were considered in \cite{Denecker95a,Denecker98c}. The
idea is to read a logic program as an inductive definition. From the
very start, logic programs were considered as {\em definitions}. This
was Clark's basic idea with the completion semantics.  Note that
Clark's completed definitions are identical to the way (non-recursive)
definitions are expressed, for example in Beth's studies.

The evident problem with Clark completion semantics is that it does
not deal well with inductive definitions. On the other hand, least
model semantics is known to deal right with positive inductive
definitions such as transitive closure. Using the above terminology,
the least model semantics can be said to be possible state semantics
for the reading of Horn programs as inductive definitions. In fact, as
pointed out in \cite{Denecker98c}, the methods that have been used to
characterize monotone induction are identical to those that were used
to characterize least model semantics of Horn programs.

Can we extend the view of Horn logic programs as inductive definitions
to programs with negation? In \cite{Denecker98c}, I have argued that
the use of induction in mathematics is not restricted to positive
induction. An example is the induction in a well-founded set. To some
extend, a generalised form of non-monotone induction have been studied
in the area of inductive definitions, the so called Iterated Inductive
Definitions \cite{Feferman70}. As I showed, this formalism is
isomorphic modulo syntactic sugar with stratified logic programs under
perfect model semantics. Further on, I have pointed to several
intolerable weaknesses of this stratified approach as a formalisation
of generalised induction and have argued that these problems are
solved by the well-founded semantics. Or, the argument there was that
the well-founded semantics is a possible state semantics of the
declarative reading of logic programs as generalised inductive
definitions\footnote{Note that well-founded semantics has been
motivated both as an atomic belief set semantics and as a possible
state semantics.}.

Consequently, logic programming has not a unique declarative reading.
The modal view and the definition view are both consistent ways of
interpreting logic programs; moreover they lead to very similar model
theories, though these theories have different roles.

The above phenomenon, the existence of different consistent
declarative views on the same formalism is a potential source of
considerable confusion. In the remaining sections, I investigate 
possible confusions.


\section{Distinguishing between declarative readings}

\subsection{Comparing declarative semantics}

The mathematical relations between the least model, models of the
completion, the perfect model, stable models and well-founded model
are understood quite well. The question I consider here is what they
tell about the relations between the two types of declarative
readings.

A naive comparison of different semantics of logic programming is
misleading. For example, the collection of stable models is known to
be a subset of the collection of models of the completion. What does
this result mean on the level of the declarative readings underlying
both semantics?

Not much it seems. E.g. the possible world model of Gelfond's
embedding and Marek and Truszczynski's embedding of the program $\{ p
\la not\ q\}$ is the set $\{ \{ p \}, \{ q,p\}\}$; this is a proper
superset of the (singleton) set of models of the
completion. Consequently, in the case of this particular program, the
completion is strictly stronger than the default or AEL reading.

\subsection{Expressing knowledge {\em declaratively}}

Consider a logic with an atomic belief set semantics for its
declarative reading.  Assume that the models of a logic theory are
exactly the possible states of the world according to the expert's
knowledge. This theory {\em encodes} the expert's knowledge but
obviously, it is not necessarily a declarative representation of the
expert's knowledge. In general the declarative reading of the theory
does not justify that the models are the only possible states of the
world. In fact, it may well be that part of the axioms of the theory
are {\em false} statements about the domain of discourse.

The above phenomenon can be illustrated in LP. Since stable semantics
is an extension of least model semantics, it is suitable to encode
positive inductive definitions such as the one of transitive closure:
\[\begin{array}{l}
  p(a,a)\la\\
  p(b,c)\la\\
  tr(X,Y) \la p(X,Y)\\
  tr(X,Y) \la p(X,Z), tr(Z,Y)
\end{array}\]
The unique stable model indeed represents the unique possible state of
the graph and its transitive closure. 

But this does not necessarily mean that inductive definitions can be
expressed under the default reading of logic programs.  For example,
the meaning of this program under Gelfond's embedding is identical to
its classical logic meaning (since AEL is a conservative extension of
classical propositional logic).

This problem gives rise to unsoundness. For example, with the
inductive definition, the expert knows that $p(a,b)$ and $tr(a,b)$ are
false.  Yet, the AEL embedding entails $\neg K p(a,b)$ and $\neg K
tr(a,b)$.

In the case of the transitive closure, the modal declarative reading
(as expressed under the above mentioned AEL and DL embeddings) of the
axioms is {\em true} but too weak to justify the unique model. An
example where the declarative reading would be plainly {\em false} can
be given by a variant of the $dead$ and $alive$ example. Assume that
the expert wants to represent the definition that $dead$ means not
$alive$, as represented by $dead \lra \neg alive$. A possible way to
do this using stable semantics is:
\[\begin{array}{l}
        dead \la not\ alive\\
        alive \la not\ alive^*\\
        alive^* \la not\ alive
\end{array}\]
The two stable models (after projection on the two atoms $alive, dead$
are identical to the models of the equivalence. Yet, it is easy to see
that both the AEL and DL embedding assigns {\em false} meaning to the
first axiom. Indeed, it is not true that $dead$ is true if one does
not know that $alive$ is true.

\subsection{Mixing different declarative readings}

The presence of multiple declarative readings and multiple roles of
models raises complications on the level of methodology. This becomes
obvious when different roles are {\em mixed}.

Consider what happens if the different sorts of declarative readings
are used in the same program. Reconsider the example of the
distrustful man. Assume that we want to add the definition that to be
$happy$ means not to be $unhappy$. According to the completion, this
knowledge is correctly represented by the rule:
\[   happy \la not\ unhappy\]
What happens if we combine this rule with the reflective knowledge of
the jealous husband. Consider the program:
\[  \left \{ \begin{array}{l}
        unhappy \la not\ wife\_faithful\\
        happy \la not\ unhappy
        \end{array} \right\}
\]
This program consists now of two isomorphic statements which are both
{\em true}, but under different declarative readings. Note that again
all semantics for this program coincide. The model is $\{unhappy\}$. What
does this model mean? Consider the two options that arose earlier:
\begin{itemize}
\item
If this model is to be interpreted that $unhappy$ is true and
$wife\_faithful$ and $happy$ are false, then there is a mismatch with
our understanding because neither the truth of $unhappy$ nor even the
knowledge that $unhappy$ is true is a sufficient condition for happy
being false. I.e. $happy$ should be unknown.
\item On the other hand, if the model is interpreted as a belief set,
there is again mismatch with our understanding because since
$wife\_faithful$ is unknown, then also $unhappy$ should be unknown.
\end{itemize}
Whether the model is interpreted as a possible state or as the belief
set, it contains an error.

\section{Extending Logic Programming}

This paper focuses on the original logic programming formalism with
only the original negation as failure, and does not investigate its
extensions with classical or strong negation.  At this point, it is
clear that there is natural reason for that. A point of this paper
is exactly that in one commonsense view on logic programming, negation
as failure {\em is} classical negation. It is clear then than in this
view, the formalism cannot be further extended with another classical
or strong negation.  Extensions of logic programming with classical
negation (including disjunctive logic programming) make sense only in
the modal negation view on logic programs.

Logic programming extensions were designed as a way to solve a number
of serious disadvantages of logic programming for knowledge
representation. However, as argued in \cite{Denecker95a}, the analysis
of what are these disadvantages exactly, depends on the declarative
reading one takes. Different declarative readings lead to different
conclusions and more importantly they subsequently lead to different
ways to different extensions of the formalism. 

In the case of the modal views, the problem was the absence of
classical negation. However, as argued in \cite{Denecker95a}, the
problem of logic programming under the definition view is that a logic
program contains definitions for all predicates. For this reason, one
should extend the formalism with the possibility of leaving certain
predicates open. This view is further explored in \cite{Denecker2000c}.


\begin{thebibliography}{}

\bibitem[\protect\citeauthoryear{Apt, Blair, \& Walker}{1988}]{Apt88}
Apt, K.; Blair, H.; and Walker, A.
\newblock 1988.
\newblock {Towards a theory of Declarative Knowledge}.
\newblock In Minker, J., ed., {\em Foundations of Deductive Databases and Logic
  Programming}. Morgan Kaufmann.

\bibitem[\protect\citeauthoryear{Baral \& Gelfond}{1994}]{Baral94a}
Baral, C., and Gelfond, M.
\newblock 1994.
\newblock Logic programming and knowledge representation.
\newblock {\em Journal of Logic Programming} 19/20:73--148.

\bibitem[\protect\citeauthoryear{Clark}{1978}]{Clark78}
Clark, K.
\newblock 1978.
\newblock Negation as failure.
\newblock In Gallaire, H., and Minker, J., eds., {\em Logic and Databases}.
  Plenum Press.
\newblock  293--322.

\bibitem[\protect\citeauthoryear{Denecker}{1995}]{Denecker95a}
Denecker, M.
\newblock 1995.
\newblock {A Terminological Interpretation of (Abductive) Logic Programming}.
\newblock In Marek, V.; Nerode, A.; and Truszczynski, M., eds., {\em
  International Conference on Logic Programming and Nonmonotonic Reasoning},
  Lecture notes in Artificial Intelligence 928,  15--29.
\newblock Springer.

\bibitem[\protect\citeauthoryear{Denecker}{1998}]{Denecker98c}
Denecker, M.
\newblock 1998.
\newblock The well-founded semantics is the principle of inductive definition.
\newblock In Dix, J.; {n}as~del Cerro, L.~F.; and Furbach, U., eds., {\em
  Logics in Artificial Intelligence},  1--16.
\newblock Schloss Daghstull: Springer-Verlag, Lecture notes in Artificial
  Intelligence 1489.

\bibitem[\protect\citeauthoryear{Denecker}{2000}]{Denecker2000c}
Denecker, M.
\newblock 2000.
\newblock Extending classical logic with inductive definitions.
\newblock In Baral, C., and Truszczynski, M., eds., {\em {8th Intl. Workshop on
  Non-Monotonic Reasoning (NMR2000)}},  1--15.

\bibitem[\protect\citeauthoryear{Feferman}{1970}]{Feferman70}
Feferman, S.
\newblock 1970.
\newblock {Formal theories for transfinite iterations of generalised inductive
  definitions and some subsystems of analysis}.
\newblock In Kino, A.; Myhill, J.; and Vesley, R., eds., {\em {Intuitionism and
  Proof theory}}. North Holland.
\newblock  303--326.

\bibitem[\protect\citeauthoryear{Gelfond \& Lifschitz}{1988}]{Gelfond88}
Gelfond, M., and Lifschitz, V.
\newblock 1988.
\newblock {The stable model semantics for logic programming}.
\newblock In {\em Proc. of the International Joint Conference and Symposium on
  Logic Programming},  1070--1080.
\newblock IEEE.

\bibitem[\protect\citeauthoryear{Gelfond}{1987}]{Gelfond87}
Gelfond, M.
\newblock 1987.
\newblock {On Stratified Autoepistemic Theories}.
\newblock In {\em Proc. of AAAI87},  207--211.
\newblock Morgan Kaufman.

\bibitem[\protect\citeauthoryear{Marek \& Truszczy\'{n}ski}{1989}]{Marek89}
Marek, V., and Truszczy\'{n}ski, M.
\newblock 1989.
\newblock Stable semantics for logic programs and default reasoning.
\newblock In E.Lust, and Overbeek, R., eds., {\em Proc. of the North American
  Conference on Logic Programming and Non-monotonic Reasoning},  243--257.

\bibitem[\protect\citeauthoryear{Marek \& Truszczy\'{n}ski}{1993}]{Marek93b}
Marek, V., and Truszczy\'{n}ski, M.
\newblock 1993.
\newblock {\em Nonmonotonic Logic\ Context-Dependent Reasoning}.
\newblock Springer-Verlag.

\bibitem[\protect\citeauthoryear{Moore}{1984}]{Moore84}
Moore, R.
\newblock 1984.
\newblock {Possible-world semantics for autoepistemic logic}.
\newblock In {\em Non-Monotonic Reasoning Workshop},  344--354.

\bibitem[\protect\citeauthoryear{Przymusinski}{1989}]{Przymusinski89}
Przymusinski, T.
\newblock 1989.
\newblock On the declarative and procedural semantics of logic programs.
\newblock {\em Journal of Automated Reasoning} 5:167--205.

\bibitem[\protect\citeauthoryear{van Emden \& Kowalski}{1976}]{vanEmden76}
van Emden, M., and Kowalski, R.
\newblock 1976.
\newblock {The semantics of Predicate Logic as a Programming Language}.
\newblock {\em Journal of the ACM} 4(4):733--742.

\bibitem[\protect\citeauthoryear{{Van Gelder}, Ross, \&
  Schlipf}{1991}]{VanGelder91}
{Van Gelder}, A.; Ross, K.; and Schlipf, J.
\newblock 1991.
\newblock {The Well-Founded Semantics for General Logic Programs}.
\newblock {\em Journal of the ACM} 38(3):620--650.

\end{thebibliography}
\end{document}